\documentclass[prl,aps,epsf,twocolumn,showpacs]{revtex4}

\usepackage{amssymb,epsfig}

\usepackage{psfrag}

\psfrag{p}{$p$}  
\psfrag{pp}{$p'$}
\psfrag{q}{$q$}
\psfrag{qp}{$q'$}
\psfrag{pip}{$\pi^+$}
\psfrag{pim}{$\pi^-$}
\psfrag{pr}{$P$}
\psfrag{apr}{$\bar P$}
\psfrag{g}{$\gamma$}
\psfrag{gs}{$\gamma^*$}
\psfrag{u}{$u$}
\psfrag{db}{$\bar d$}
\psfrag{ep}{$e^+$}
\psfrag{em}{$e^-$}
\psfrag{d}{$d$} 
\psfrag{a}{$a$}
\psfrag{b}{$b$}

\newcommand{\out}{\raise-3pt\hbox{\scriptsize    out}}

\begin{document}

\title{Hadron  annihilation into
two photons and
backward  VCS in the scaling regime of QCD.}
\author{  B. Pire$^1$ and  L. Szymanowski$^{2,3}$  } 
\affiliation{$^1$CPhT,
\'Ecole  Polytechnique,  F-91128  Palaiseau,  France  \\  
$^2$  Soltan
Institute  for   Nuclear  Studies,  Warsaw,   Poland  \\  
$^3$Physique
Th\'eorique  Fondamentale,  Universit\'e  de Li\`ege,  B4000  Li\`ege,
Belgium}
%
%
\preprint{CPHT-XXX,}
\begin{abstract}   {We  study   the   scaling  regime   of  hadron   -
(anti)-hadron  annihilation into a  deeply virtual  photon and  a real
photon,  $H \bar  H \to  \gamma^*\gamma $,  and   backward virtual
Compton scattering, $\gamma^* H \to H  \gamma $. We advocate that there is
a kinematical region where  the scattering amplitude factorizes into a
short-distance matrix element and a long-distance dominated object : a
transition distribution amplitude  which describes the hadron to
photon transition.}
\end{abstract}
%
\maketitle


\noindent {\bf 1. Introduction}

\noindent There now exists  a successful description of deep exclusive
reactions  in  terms   of  distribution  amplitudes \cite{ERBL}  and/or
generalized  parton distributions \cite{DiehlPR}  on the  one  side and
perturbatively   calculable  coefficient  functions   describing  hard
scattering at  the partonic  level on the  other side.  The pioneering
papers \cite{ERBL}  and  \cite{Dvcs} have  opened  the  way  to a  real
understanding of many processes  that share with hadronic form factors
and  deeply virtual Compton  scattering (dVCS)  in the  nearly forward
region    some    basic    properties    and   in    particular    the
factorization \cite{Collins}  of  the  long distance  dominated  matrix
element  of the  non-local correlator  of  two quark  or gluon  fields
between two hadronic states. Many  processes have been studied in this
framework  \cite{meson}.  Cases  have   been  studied  where  the  
generalized parton distribution (GPD)
describes  the   transition  of   a  nucleon  to   a  nucleon,   to  a
resonance \cite{NDelta}, a  nucleon-meson continuum state \cite{NucMes}
or  the  exotic pentaquark \cite{DPS}.   We  consider  here a  somewhat
different class of reaction, where the t-channel exchange carries most
of the  quantum numbers of the  hadron. The simplest  examples are the
annihilation:
    \begin{equation} 
\bar{p} p \to  \gamma ^* \gamma \to e^+e^- \gamma
  \label{pbarp}
\end{equation} 
in the  near forward region and large virtual photon invariant mass 
$Q$ which will  be studied in
detail  at GSI \cite{Panda},  and  the VCS  reaction  in the  backward
region (i.e. when the real photon is emitted in the direction of the 
proton momentum in the CM system)
\begin{equation} 
\gamma ^* p \to p \gamma \:,
\label{bdVCS}
\end{equation} 
which  is currently under  experimental investigation
at JLab \cite{JLab}.
 
 The crucial  observation that we want  to make is  that the amplitude
for  the process  (\ref{pbarp})  is  quite analogous  to  the one  for
process (\ref{FF})
 \begin{equation}  \bar{p} p  \to  \gamma ^*  \to e^+e^-  ~~~~~~~~~~~~
\gamma ^* p \to p\;,
\label{FF}
\end{equation}  which are  controlled  by the  timelike and  spacelike
proton form factors, respectively, the asymptotic form of which may be
calculated in QCD \cite{ERBL}, through  the factorization of proton and
antiproton  distribution amplitudes  on  the one  hand,  and the  hard
processes $\bar q \bar q \bar q q q q \to \gamma^*$ or $\gamma^* q q q
\to q q  q$ on the other  hand.  The only change needed  is to replace
the proton distribution amplitude which is defined from the 
correlator
$ \langle 0|\, {u}(z_{1})\, {u}(z_{2})\, {d}(z_{3}) \,|p \rangle$ by a
quantity derived from a slightly  different matrix element of the same
operator,  {\em i.e.} the  correlator $\langle  \gamma|\, {u}(z_{1})\,
{u}(z_{2})\, {d}(z_{3}) \,|p \rangle$. This symbolic notation omits the 
gauge links
between quarks and their spin and  colour indices.

\begin{figure}[t]
\centerline{\epsfxsize8.5cm\epsffile{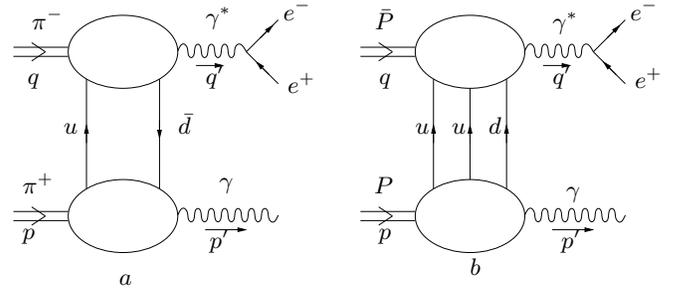}}
\caption[]{\small
The factorization of the annihilation process $\bar H \;H\to \gamma^*
\,\gamma$ into a hard subprocess (upper blob) and a transition
distribution
amplitude (lower blob) for the meson case (a) and the baryon case (b).
 }
\label{fig:1}
\end{figure}

This is what we define, see Fig.~1, as the hadron to photon transition 
distribution
amplitude (TDA), which enters in
the  expression of  the  amplitude for  process  (\ref{pbarp}) in  the
kinematical region where hard and soft processes decouple, namely when
the real  photon momentum is  almost collinear to the  incoming 
($\pi^+$ or $p$ in Fig.~1) hadron
 momentum as in the case of QCD factorization in dVCS. An other useful 
reaction of reference is the reaction $\pi^- \;p\,\to\,\gamma^*\,n$ 
\cite{TCS} which has been analysed in the framework of QCD 
factorization. If we restrict this reaction to a meson target, the 
obtained reaction $\pi^-\,\pi^+ \, \to \, \gamma^*\,\pi^0$ looks much the 
same as the one we consider here. 
One can apply to this last process the proof of factorization given in
Ref. \cite{Collins} for the vector meson production \footnote{We 
thank J. Collins for useful 
correspondence on this
point.}.


\noindent
Such an extension of the  factorization proof has already been briefly
advocated  in  Ref. \cite{FPPS} for  the  electroproduction of  leading
baryons or antibaryons. As was advocated there, factorization in hard 
exclusive  processes such as $\gamma_L^* p  \to \rho p'$ is basically due 
to the fact that the meson originates from a small size $q \bar q$ object
which  is  generated  by  a  hard  scattering  (the  hardness  of  the
scattering controlling the  initial size of the color  singlet). In the
case under study,  the only difference is that the small  size $q q q$
object carries baryon number.

\noindent
In this  letter, we  shall restrict to  the case of  spinless hadrons,
and study the reactions
\begin{equation}
\pi^-\,\pi^+ \to \gamma^* \, \gamma \,,\;\;\;\;\; \gamma^*\, \pi^+ \to 
\pi^+ \gamma\;,
\end{equation}
 with  a few  qualitative comments  on the
extension of this analysis  to the phenomenologically more interesting
proton case, which will be studied in detail in a future work. We also
restrict to the longitudinal polarization of the virtual photon.
 
\vspace{0.3cm}
\noindent  {\bf   2.  The  $H  \to   \gamma$  transition  distribution
amplitude}

\noindent Let  us take  a closer look  at the  transition distribution
amplitudes  that occur  in the  processes we  are interested  in.  For
their definition we introduce light-cone coordinates $v^\pm = (v^0 \pm
v^3) /\sqrt{2}$ and  transverse components $v_T = (v^1,  v^2)$ for any
four-vector $v$.   The skewness variable  $\xi = (p-p')^+  /2P^+$ with
$P=(p+p')/2$  describes  the loss  of  plus-momentum  of the  incident
hadron.  The momentum transfer is $\Delta=p'-p$.

We define the corresponding $H \to \gamma$ leading twist TDAs in the 
mesonic case as ($\varepsilon$ is the polarization of a photon) 
\begin{eqnarray}
 \label{V} & \int \frac{d z^-}{2\pi}\, e^{ix P^+ z^-} 
\langle     \gamma(p',\varepsilon)|\, O_V^\mu
\,|\pi^+(p) \rangle \Big|_{z^+=0,\,  z_T=0} \nonumber \\ 
&=\frac{1}{P^+} \frac{i\;e}{f_\pi}\epsilon^{\mu \nu \rho 
\sigma}\varepsilon_{\perp\nu}
P_\rho \Delta_{\perp \sigma}\;V(x,\xi,t) 
\nonumber \\
& O_V^\mu = \bar{d}(-z/2)\,\gamma^\mu
\,{u}(z/2)
 \end{eqnarray} 
\noindent
\begin{eqnarray}
  \label{A} &\int \frac{d z^-}{2\pi}\, e^{ix P^+z^-}       
\langle       \gamma(p',\varepsilon)|\,  O_A^\mu   \,|\pi^+(p)    \rangle 
\Big|_{z^+=0,\,  z_T=0}   
\nonumber \\ 
&= \frac{1}{P^+} \frac{e}{f_\pi} (\vec 
\varepsilon \cdot \vec \Delta) P^\mu \;A(x,\xi,t)
\nonumber \\
& O_A^\mu = \bar{d}(-z/2)\,\gamma^\mu 
\gamma_{5}\,{u}(z/2) 
 \end{eqnarray}
\begin{eqnarray}
  \label{T} &\int \frac{d z^-}{2\pi}\, e^{ix P^+z^-}
\langle       \gamma(p',\varepsilon)|\, O_T^{\mu\nu}  \,|\pi^+(p)    
\rangle \Big|_{z^+=0,\,  z_T=0}  \nonumber \\ 
&= \frac{e}{P^+} \epsilon^{\mu \nu \rho  
\sigma} P_\sigma \left[ 
 \varepsilon_{\perp \rho} 
 T_1(x,\xi,t) - \frac{1}{f_\pi} (\vec \varepsilon \cdot \vec 
\Delta)   \Delta_{\perp \rho} T_2(x,\xi,t) \right]  \nonumber \\
& O_T^{\mu \nu} = \bar{d}(-z/2)\,
\sigma^{\mu \nu} \,{u}(z/2)\;,   
\end{eqnarray}
where the first two TDAs, $V(x,\xi,t)$ and $A(x,\xi,t)$ are chiral even 
and the latter ones, $T_i(x,\xi,t), i=1,2$, are chiral 
odd. In definitions (\ref{V}), (\ref{A}) and (\ref{T}) it is assumed 
that like in the usual GPDs a Wilson link along the light-cone is included 
to ensure the QCD-gauge invariance for non local operators. We omit this 
link to simplify the notation.
 $f_\pi$ is the pion decay constant. The four leading twist
TDAs are linear combinations of the four independent helicity amplitudes 
for the process $q \pi \to q \gamma$.

 As in the  case of GPDs, the signs of  $x \pm \xi$ determine the partonic
interpretation  of   the  TDAs.   One  thus
distinguishes  in complete analogy  with usual  GPDs the  DGLAP 
(Dokshitzer-Gribov-Lipatov-Altarelli-Parisi) region
where $x+\xi$  and $x-\xi$ have the same  sign, where the
physics  resemble the usual  parton distribution  case, from  the ERBL 
(Efremov-Radyushkin-Brodsky-Lepage)
region for which  $-\xi < x  < \xi$ and  where a quark-antiquark pair is 
extracted from
the meson state.  
  The interpretation of TDAs is  more explicit when they are expressed
as the  overlap of  light-cone wave functions  for the hadron  and the
photon.    Let
us emphasize  that all possible  spectator configurations have
to be summed over in  the wave function overlap, including Fock states
with additional partons in the hadron and in the photon.

\vspace{0.3cm}

Interesting  sum rules  may  be derived  for  the  meson to  photon
TDAs. Since the local matrix element  describes - up to a charge factor
- the  photon to neutral  meson transition form factor,  measurable in
$\gamma \gamma ^* $ collisions, we obtain
 
\begin{equation} \int^1_0 dx \left( Q_u\,V^u(x,\xi, t) 
- Q_d\,V^d(x,\xi,t)\right)= \sqrt{2}\,f_\pi\,F_{\pi \gamma} (t)\,,
\label{srV}
\end{equation}
where $Q_u=2/3$, $Q_d=-1/3$ and $V^q(x,\xi, t)$ is the TDA related to the 
operator built from the quark $q$, see Eq.~(\ref{V}).
Current algebra fixes the value of the right hand side at $t=0$ since 
$F_{\pi \gamma} (t=0)= \frac{\sqrt{2}}{4\pi^2 f_\pi}, 
f_\pi=133\;$MeV \cite{BL81}. 

In an analogous way one can derive the sum rule for the axial TDAs, related to the 
axial-vector form-factors measured in the weak decays of mesons 
$\pi^+ \to l^+ \, \nu_l \, \gamma$ \cite{weak}.
We get
\begin{equation} \int^1_0 dx A(x,\xi,t)= f_\pi\,F_A(t).
\label{srA}
\end{equation}
 These two sum rules allow to constrain possible parametrizations of the TDAs.
Note, in particular, the $\xi$ independence of both relations. 
On the other hand we do not know any sum rule constraining chirally-odd
TDAs.

\vspace{0.3cm}
\noindent  QCD  radiative corrections  lead  as  usual to  logarithmic
scaling violations.  The scale dependence of the meson to photon TDAs
is governed by evolution equations which are  the
evolution equations for usual non-singlet GPDs \cite{Dvcs}, 
\begin{equation}  
\mu^2\,\frac{d}{d\,\mu^2} A(x,\xi,t)= \int\limits_{-1}^1\,dx'
\frac{1}{\xi}\,V_{NS}\left( \frac{x}{\xi},\frac{x'}{\xi} \right) 
A(x',\xi,t) \;,
\label{eveq}
\end{equation} 
where to the leading order accuracy
\begin{eqnarray}
&&V_{NS}(x,x') = 
\frac{\alpha_s}{4\,\pi}C_F \left[\rho(x,x')\left[ 
\frac{1+x}{1+x'}\left( 1+\frac{2}{x'-x}  \right) \right] \right. 
\nonumber \\
&&\left. + [x\to -x, x' 
\to -x']  \right]_+ \nonumber \\
&& \rho(x,x')=\theta(x'\geq x \geq -1) - \theta(x' \leq x \leq -1)\;,
\label{kernel}
\end{eqnarray} 
with $C_F=(N_c^2-1)/(2N_c)$,
 and the same equations are satisfied by the 
other leading order TDAs, 
$V(x,\xi,t)$ and  $T_i(x,\xi,t),i=1,2$.

\vspace{0.3cm}
\noindent  {\bf 3.  $ \pi^- \, \pi^+  \to l^+l^-  \gamma $  in  the 
scaling regime}

\noindent Let  us now study  the annihilation channel  $ \pi^-\,\pi^+ \to
l^+l^-\gamma $ in the kinematical region where
$Q^2, s \to \infty$ with $x_B = Q^2/s$ fixed. $Q^2$ is the invariant 
mass of the lepton pair and $s$ is the center of mass energy squared. 

\noindent  The  amplitude  is   the  sum  of  two  contributions.  The
Bremsstrahlung process where the real  photon is radiated from a final
lepton yields  an amplitude proportional to the  timelike hadron form
factors  $F_{\pi}(s)$. Such  an amplitude  is strongly  peaked  when the
photon  is  collinear to  one  of the  final  leptons.  It should  not
interfere  much with  the process  we are  mostly interested  in which
favors  the kinematics  where the  photon  is mostly  parallel to  the
$\pi^+$ \cite{rem}. 

\noindent
The $\pi^- \,\pi^+ \to \gamma_L^* \gamma $ contribution yields an 
amplitude
which is proportional to the TDAs we have defined with $\xi$ connected
with $x_B$ by $\xi \approx \frac{x_B}{2-x_B}$ in the Bjorken
limit. It reads
\begin{equation}
\label{amp} 
{\cal  M} (Q^2, \xi)=  \int dx dz \phi(z)  M_{h}(z,x,\xi) A(x, \xi, t)\;,
\end{equation}
 where the hard amplitude is
\begin{eqnarray}
&&M_{h}(z,x,\xi) =  \\ 
&&\frac{4\,\pi^2\,\alpha_{em}\,\alpha_s\,C_F}{N_C\,Q}
\frac{1}{z\,\bar z}\left(\frac{Q_u}{x-\xi-i\epsilon} + 
\frac{Q_d}{x+\xi+i\epsilon}  \right)\vec \varepsilon \cdot \vec \Delta\,,
\nonumber
\end{eqnarray}   
and  $\phi(z)$   is  the   meson  distribution
amplitude, $\bar  z = 1-z$.

\noindent The  scaling law for  the amplitude  is 
\begin{equation} 
{\cal M}(Q^2, \xi) \sim \frac{\alpha_s(Q^2)}{Q}\;,
\label{scaling}
\end{equation} 
up to  the logarithmic corrections due to DA and TDA anomalous dimensions. 
 This behaviour is related to the one of
the electromagnetic hadron form factor.

\vspace{0.3cm}
\noindent {\bf 4.  Backward VCS}

\noindent The backward VCS reaction $ e  H \to e' H \gamma $ when the
final photon flies approximatively in the direction of the initial 
hadron receives  contributions  from two  competing  subprocesses, namely  
the
Bethe-Heitler  process where  the  final photon  is  emitted from  the
lepton line and the hadronic dVCS process where $\gamma$ is emitted from 
the
hadron.    Contrarily  to  the   case  of   near  forward   dVCS,  the
Bethe-Heitler   process  is  strongly   suppressed  in   the  backward
kinematics. One may thus quite  safely ignore its contribution in this
study \cite{rem}.

 The hadronic  backward VCS  process is the  spacelike analog  of the
$\bar H  H \to \gamma_L^* \gamma  $ process discussed  in the previous
section.  The  correspondence  between  these two  amplitudes  may  be
inferred, via crossing, from the  analysis of Ref.~\cite{TCS}: at  Born 
level and to
leading twist, one obtains the amplitudes for backward dVCS from those
of the reaction $\bar H H \to \gamma_L^* \gamma $ by changing the sign
of  the imaginary  part  and  reversing the  beam  and virtual  photon
polarizations. To this accuracy, both processes thus carry exactly the
same information  on the  TDAs. One should however remember that 
already  at the
form  factor  level,  there  is  an  experimental  difference  between
timelike and spacelike exclusive quantities which may be the signature
of higher twist or higher order corrections \cite{tvs}.

The scaling law for the amplitude is the same as the one written above
in Eq. (\ref{scaling}).

\vspace{0.3cm}
\noindent {\bf 5. Conclusions}

\noindent 
We have defined the new transition distribution amplitudes 
$H \to \gamma $, i.e. which parametrize the matrix elements of 
light-cone operators between  very different initial and final states; 
this  generalizes the concept of GPDs for non-diagonal transitions.
In our analysis we assumed that the momentum transfer square $t$ is small, 
$Q^2$ is large and leads to the scaling.
In a real experiment, at finite (and not so large ) $Q^2$  and when $-t$
 increases from zero to the order of $Q^2$,  the real photon has to be
 treated
 as a part of  the hard physics and the  process becomes a fixed angle
 process. Such case  may be discussed along the lines of
  Ref.~\cite{ERBL} or analysed with the use of the handbag mechanism 
of Ref.~\cite{ra}.

The TDAs we introduced are in particular relevant for the
 reactions 
involving baryon number exchanges. 
We  have discussed  two 
new cases  of factorization  in hard
exclusive reactions, the soft parts of which are described by the same
hadron to photon TDA. 
Although, for simplicity, our study was concentrated on the 
mesonic case, we stress that most of our conclusions apply also for 
the phenomenologically more interesting, but also technically more 
difficult, baryonic case. 

Let us emphasize the main differences between the mesonic and the baryonic
cases: 

\noindent
-  The relevant matrix element involves three quark fields
$   \langle \gamma|\, {u}^{\alpha}(z_{1})\, 
[z_1;z_0]\,{u}^{\beta}(z_{2})\, [z_2;z_0]\,
   {d}^{\gamma}(z_{3}\,[z_3;z_0]) \,|p \rangle \Big|_{z_{i}^+=0,\, 
{z}_{i}^T=0}$ and consequently it will involve more independent spinor 
structures than in the meson case. 

\noindent
- The number of independent helicity amplitudes for the
process $p \to  qqq \gamma$ is 16 (there are in this case $2^5$
helicity  amplitudes but  parity  conservation  relates  amplitudes
differing by all the signs of helicities). This allows to predict that 
there will be 16 independent, leading twist TDAs for the $p \to \gamma$ 
transition. As in the meson case, where only one out of four TDAs actually 
enters Eq.~(\ref{amp}), less than 16 TDAs are expected to contribute to 
the amplitude of the processes under study.

\noindent
- A slight generalization of the
scaling law written in
Eq.~(\ref{scaling}) dictates the high $Q^2$ behaviour of the baryonic
amplitudes, up to logarithmic corrections
\begin{equation}
{\cal M}(Q^2, \xi) \sim Q^{-2n_{v}+3}  \;,
\label{scalingb}  
\end{equation}
where $n_{v}$  is the number of valence  partons in the
hadron ($n_v=3$ for the proton).

\noindent
- The QCD evolution equations for the TDAs in the baryonic case are more 
involved since  the radiated gluon may be coupled to 
each pair of exchanged quarks and the  evolution kernel depends  on the
direction  of   these  quark  momenta. 
 New  evolution equations may  be written in a way which takes  care of
the different sectors in longitudinal momentum fractions of the three 
exchanged quarks.


\noindent
 The  framework which we present in the present paper 
is  very different
from  the handbag  dominance model \cite{ra}  which has been proposed 
to describe the annihilation reaction $p \bar p \to \gamma \gamma$.
  Such  a  model, based on an analogy with the reaction $\gamma^*\,\gamma 
\to \pi^+\,\pi^-$ at small energies \cite{DGPT},  
  predicts  a  different
scaling behaviour  at large $Q^2$ and a  different helicity structure.
Although these two approaches deal with different kinematical domains
(the present one studies the forward and backward cases, those of 
Ref.~\cite{ra} 
concentrates on fixed large angle kinematics) which are well distinct at 
asymptotic energies,  actual experimental data at 
moderate values of $Q^2$ and $s$ may not well separate these two 
regions.
It should be important to investigate in detail the phenomenological 
differences 
of these two approaches.
Experimental  tests of  our picture  should  be feasible  in the  near
future.
Already now experimental data exist on backward VCS at Jlab up to 
$Q^2 = 1$
GeV$^2$ \cite{JLab}. 
Further experiments on backward VCS are planned with higher
energy and rates at JLab, Hermes and Compass experiments.
The Panda and PAX experiment  with  the
proposed $1.5-15 $ GeV  high-luminosity antiproton storage ring (HESR)
at GSI \cite{Panda} intend to study the proton
antiproton annihillation reaction. 
 Let us finally note that $p \to \pi$ TDAs may also be defined in an 
analogous way \cite{PS3}, allowing a similar description of reactions such as
\[
p \bar p \to \gamma^*\;\pi \;\;\;\mbox{or}\;\;\;\;\gamma^*\;p\to 
\,p\;\pi\;.
\]
Again these processes are likely to be measurable in forthcoming 
experiments at GSI, JLab and HERMES.

\vspace{0.3cm}
\noindent {\bf Acknowledgments.}

\noindent  We acknowledge useful  discussions and  correspondence with
V.~Braun,  M.~Diehl,  M.~D\"  uren,  M.~Guidal,  G.~Korchemsky,
L.~{\/L}ukaszuk,
M.~Polyakov, J.P.~Ralston, C.~Roiesnel, O.V.~Teryaev and S.~Wycech.   
This  work  is  supported  by  the
French-Polish scientific agreement Polonium, the Joint Research
Activity "Generalised Parton Distributions" of the european I3 program
Hadronic Physics, contract RII3-CT-2004-506078 and the Polish Grant 
1 P03B 028 28. 
L.Sz. is a Visiting Fellow of 
the Fonds National pour la Recherche Scientifique (Belgium).



\begin{thebibliography}{99}
\bibitem{rem} Detailed study will be carried out in the phenomenologically 
interesting cases in a future work.
\end{thebibliography}
\end{document}